\documentclass[aps,prd,twocolumn,preprintnumbers,amsmath,amssymb,
superscriptaddress,showpacs,floatfix]{revtex4}

\usepackage{graphicx}
\usepackage{dcolumn}
\usepackage{bm}
\usepackage{epsfig}
\usepackage{graphics}
\usepackage{latexsym}

\def\beq{\begin{equation}}
\def\eeq{\end{equation}}
\def\bea{\begin{eqnarray}}
\def\eea{\end{eqnarray}}
\newcommand{\beqs}{\begin{subequations}}
\newcommand{\eeqs}{\end{subequations}}

\newcommand{\cref}[1]{Ref.~\cite{#1}}

\newcommand{\vev}[1]{\left<#1\right>}

\newcommand{\hh}{{\ensuremath{I{\kern-2.6pt h}}}}
\newcommand{\bhh}{{\ensuremath{\bar{I{\kern-2.6pt h}}}}}

\begin{document}

\preprint{UT-STPD-16/01}

\title{Diphoton resonances in a $U(1)_{B-L}$ extension of the minimal \\
supersymmetric standard model}

\author{G. Lazarides}
\email{lazaride@eng.auth.gr} \affiliation{School of Electrical and
Computer Engineering, Faculty of Engineering, Aristotle University
of Thessaloniki, Thessaloniki 54124, Greece}
\author{Q. Shafi}
\email{shafi@bartol.udel.edu}
\affiliation{Bartol Research Institute, Department of Physics and 
Astronomy, University of Delaware, Newark, DE 19716, USA}

\date{\today}

\begin{abstract}
Inspired by the 750 GeV diphoton state recently reported by ATLAS 
and CMS, we propose a $U(1)_{B-L}$ extension of the MSSM which 
predicts the existence of four spin zero resonance states that are 
degenerate in mass in the supersymmetric limit. Vector-like fields, 
a gauge singlet field, as well as the MSSM Higgsinos are 
prevented from acquiring arbitrary large masses by a $U(1)$ 
R-symmetry. Indeed, these masses can be considerably lighter than 
the $Z'$ gauge boson mass. Depending on kinematics the resonance 
states could decay into right handed neutrinos and sneutrinos, 
and/or MSSM Higgs fields and Higgsinos with total decay widths in 
the multi-GeV range.
\end{abstract}

\maketitle


The recently reported \cite{ATLAS/CMS} 750 GeV diphoton resonance 
by ATLAS and CMS, if confirmed during Run III at the LHC, would 
have far reaching ramifications in our 
quest for new physics beyond the Standard Model (SM). These 
preliminary results have triggered, not surprisingly perhaps, 
a flurry of theoretical papers \cite{theory} offering a large 
variety of 
plausible extensions of the SM in order to explain the reported 
diphoton excess.

In this paper we propose an 
extension 
of the Minimal Supersymmetric Standard Model (MSSM) which 
naturally yields resonance states in the TeV range. A simple 
implementation of this framework is realized in a local $U(1)_{B-L}$ 
extension of the MSSM gauge symmetry. In contrast to radiative 
electroweak breaking implemented in the MSSM, the additional 
local symmetry in our case is spontaneously broken at tree level 
with a superpotential $W$ whose form is uniquely determined by 
a combination of the underlying gauge symmetry and a $U(1)$ global 
R-symmetry. The construction of $W$ utilizes an appropriate pair 
of Higgs superfields ($\Phi$, $\bar{\Phi}$) as well as a gauge singlet 
superfield $S$. The resonance states  
arise from the scalar components of the $S-\Phi-\bar{\Phi}$ system, 
and their mass is determined, in the supersymmetric (SUSY) limit, by a 
dimensionless parameter in W which can be much smaller, 
if needed, compared to the typical order unity or so gauge 
coupling constant. Thus, if required, the resonance states can be 
significantly lighter than the mass of the $Z'$ gauge boson 
associated with $B-L$. Note that with global $U(1)_{B-L}$ any 
constraint arising from $Z'$ goes away. This is a plausible 
alternative to local $U(1)_{B-L}$ considered here.

The spontaneous breaking of $U(1)_{B-L}$ leaves 
SUSY unbroken. The symmetry breaking scale $M$ may be 
much larger than the TeV SUSY breaking scale. Superpotentials 
of this type have previously 
been employed by Dvali and Shafi \cite{Dvali:1994wj} in their 
construction of SUSY trinification models based on 
$SU(3)_c\times SU(3)_L\times SU(3)_R$, and later in the 
construction of SUSY hybrid inflation models 
\cite{Dvali:1994ms,Copeland:1994vg}.

The scalar component of $S$ acquires a non-zero vacuum expectation 
value (VEV) proportional to 
$m_{3/2}$ after SUSY breaking \cite{Dvali:1994wj,Dvali:1997uq}, and 
this has been utilized in the past \cite{Dvali:1997uq,King:1997ia} 
to resolve the MSSM 
$\mu$ problem. The R-symmetry also protects $S$ from acquiring 
arbitrarily large masses. In the present scheme we also use 
this $\sim{\rm TeV}$ VEV of $S$ to provide masses to suitable 
vector-like fields including colored fields which play a role 
in the production and subsequent decay of the scalar 
resonance(s). 



The renormalizable superpotential of the MSSM with R-parity possesses 
three global symmetries, namely 
the baryon number $U(1)_B$, lepton number $U(1)_L$ and a R-symmetry 
$U(1)_R$, where, for simplicity, we ignore the tiny non-perturbative 
violation of $B$ and $L$ by the $SU(2)_L$ instantons. The new local 
$U(1)$ symmetry, which we identify as $U(1)_{B-L}$, is to be 
spontaneously broken at some scale $M$, and we prefer to implement this 
breaking by a SUSY generalization of the 
Higgs mechanism. 
Motivated by the MSSM example, we 
require that the new superpotential $W$ respects the global $U(1)_B$ 
and $U(1)_L$ symmetries as well as a global $U(1)$ R-symmetry. 

The full renormalizable  superpotential is 
\bea
W &=&y_uH_{u}qu^c + y_dH_{d}qd^c + y_{\nu}H_{u}l\nu^c +y_{e}H_{d}le^c
\nonumber
\\
& & +\kappa S (\Phi\bar{\Phi}-M^2) +\lambda_{\mu} SH_{u}H_{d}+ 
\lambda_{\nu^c}\bar\Phi \nu^c\nu^c 
\nonumber
\\ 
& & +\lambda_{D} S D\bar{D}+ \lambda_{q}Dqq + \lambda_{q^c}\bar{D}u^cd^c, 
\label{W}
\eea
where $y_u$, $y_d$, $y_{\nu}$, $y_e$ are the Yukawa coupling constants 
and the family indices  are generally suppressed for simplicity. Here 
$q$, $u^c$, $d^c$, $l$, $\nu^c$, $e^c$ are the usual quark and lepton 
superfields of MSSM including the right handed neutrinos $\nu^c$, and 
$H_{u}$, $H_{d}$ are the standard electroweak Higgs superfields. The 
gauge singlet $S$ has necessarily the same R-charge as $W$, which we 
take to be 2. Consequently, $H_{u}$, $H_{d}$ have opposite R-charges, 
which can be brought to zero by an appropriate hypercharge ($Y$) 
transformation. The R-charges of $u^c$ ($\nu^c$) and $d^c$ ($e^c$) are 
equal and, thus, $B$ and $L$ transformations can make the R-charges of 
$q$, $u^c$, $d^c$, $l$, $\nu^c$, $e^c$ all equal to unity. 

In order to determine the R-charges and $B-L$ quantum numbers of the 
SM singlets $\Phi$, $\bar{\Phi}$, we introduce the coupling 
$\bar{\Phi}\nu^c\nu^c$, which implies that their $B-L$ charge is 
$2$, $-2$ respectively, and their R-charges are zero. This 
coupling generates masses for the right handed neutrinos after the 
breaking of $U(1)_{B-L}$ to its $Z_2$ subgroup by the VEVs of 
$\Phi$, $\bar{\Phi}$.

We also introduced the coupling $SD\bar{D}$, where $D$ ($\bar{D}$) are
color triplet (antitriplet) and $SU(2)_L$ singlet superfields. (Color 
triplet vector-like superfields $D$, $\bar{D}$ are perhaps best 
motivated in the framework of GUT symmetry $E_6$.) To determine 
the charges of these fields we need at least one additional coupling. 
Taking the coupling $Dqq$, which is a color and $SU(2)_L$ singlet, the 
R-charges of $D$, $\bar{D}$ vanish. Also, the $Y$ charge of $D$ becomes -1/3 
and consequently, the hypercharge of $\bar{D}$ is 1/3. Finally, the $B-L$ 
charge of $D$ is -2/3 and that of $\bar{D}$ is 2/3 with their lepton numbers 
vanishing. Note that the coupling $\bar{D}u^cd^c$ is also present since it 
respects all the symmetries of the model. It is also worth mentioning that 
the $Z_2$ subgroups of both $U(1)_R$ and $U(1)_{B-L}$ coincide with the 
$Z_2$ matter parity under which all the ordinary (anti)quark and 
(anti)lepton superfields are odd, with the rest of the superfields being
even. This symmetry remains unbroken by all the soft SUSY 
breaking terms and all the VEVs. (See Ref.~\cite{Kibble:1982ae} for a more 
general discussion of unbroken $Z_2$ from $SO(10)$.)  In 
Table~\ref{tab:fields}, we summarize all the superfields of 
the model together with their transformation properties under the SM gauge 
group ($G_{SM}=SU(3)_c\times SU(2)_L\times U(1)_Y$) and their charges under 
the global symmetries $U(1)_B$, $U(1)_L$, and $U(1)_R$.

\begin{table}[!t]
\caption{Superfield content of the model.}
\begin{tabular}{c@{\hspace{.8cm}}
c@{\hspace{.8cm}} c@{\hspace{.8cm}} c@{\hspace{.8cm}}c}
\toprule
{Superfields}&{Representions}&\multicolumn{3}{c}{Global Symmetries}
\\
{}&{under
$G_{SM}$}&{$B$} &{$L$} &{$R$}
\\\colrule
\multicolumn{5}{c}{Matter Superfields}\\\colrule
{$q$} &{$({\bf 3, 2}, 1/6)$}&$1/3$& $0$& $1$
\\
{$u^c$} & {$({\bf \bar 3, 1},-2/3)$}&$-1/3$&{$0$}&$1$
\\
{$d^c$} & {$({\bf \bar 3, 1},1/3)$}  &$-1/3$&{$0$}&$1$
\\
{$l$} &{$({\bf 1, 2}, -1/2)$} &$0$& $1$&$1$
 \\
{$\nu^c$} & {$({\bf 1, 1}, 0)$}&{$0$}&{$-1$} &$1$
\\
{$e^c$} & {$({\bf 1, 1}, 1)$}&{$0$}&{$-1$}&$1$ 
\\
\colrule
\multicolumn{5}{c}{Higgs Superfields}
\\\colrule
{$H_u$} & {$({\bf 1, 2},1/2)$} &$0$&$0$&$0$
\\
{$H_d$} & {$({\bf 1, 2},-1/2)$} &$0$&$0$&$0$
\\
\colrule
{$S$} & {$({\bf 1, 1},0)$}  &$0$&$0$&$2$ 
\\ 
{$\Phi$} &{$({\bf 1, 1},0)$} & {$0$}&{$-2$}&{$0$} 
\\
{$\bar\Phi$}&{$({\bf 1, 1},0)$}&{$0$}&{$2$}&{$0$} 
\\
\colrule
\multicolumn{5}{c}{Vector-like Diquark Superfields}
\\\colrule
$D$&{$({\bf 3, 1},-1/3)$} &$-2/3$& $0$ &$0$
\\
$\bar{D}$&{$({\bf \bar 3, 1},1/3)$}  &$2/3$&$0$& $0$
\\
\botrule
\end{tabular}\label{tab:fields}
\end{table}

The superpotential in Eq.~(\ref{W}) is
the most general renormalizable superpotential which obeys the SM gauge
symmetry and the global symmetries $U(1)_B$, $U(1)_L$, and $U(1)_R$. Had 
we removed the separate baryon and lepton number symmetries and kept only 
the gauge $U(1)_{B-L}$ symmetry, the renormalizable superpotential terms
$\bar{D}ql$, $Du^ce^c$, and $Dd^c\nu^c$ would be present leading to fast 
proton decay and other baryon and lepton number violating effects 
\cite{Lazarides:1998iq}. The spontaneous breaking of $U(1)_{B-L}$ to its 
$Z_2$ subgroup will generate a network of local cosmic strings. Their 
string tension, which is determined by the scale $M$, is relatively small 
and certainly satisfies the most stringent relevant upper bound from pulsar 
timing arrays \cite{pulsar}.     

The `bare' MSSM $\mu$ term is now replaced by a term $SH_{u}H_{d}$, so that 
the $\mu$ term is generated after $S$ acquires a non-zero VEV of order TeV 
from soft SUSY breaking \cite{Dvali:1997uq}. (In the SUSY limit the VEV of 
$S$ is zero.) The VEV of $S$ also plays an essential role, as we will see, 
in the generation of masses for the vector-like fields $D$, $\bar{D}$ that 
are crucial in the production and decay of the diphoton resonance(s).

The spontaneous breaking of $U(1)_{B-L}$ implemented with the fields 
$S$, $\Phi$, $\bar{\Phi}$ delivers, in the exact SUSY limit, four spin 
zero particles all with the same mass given by $\sqrt{2}\kappa M$.
This mass, even for $M\gg 1~{\rm TeV}$, can be of 
order TeV (more precisely $\simeq 750~{\rm GeV}$ in the present case) by 
selecting an appropriate value for $\kappa$. We should point out though 
that depending on the SUSY breaking mechanism, the four resonance states 
may end up with significantly different masses. The VEV of $S$, with 
suitable choice of the gauge and R-charges, yields 
masses for the appropriate fields that are vector-like under the MSSM 
gauge symmetry. This is in addition to possibly additional 
such fields that acquire masses from their coupling to the Higgs fields 
with VEV $M$ that break the $U(1)_{B-L}$ gauge symmetry. The diquarks 
associated with the vector-like fields may be found \cite{Gogoladze:2010xd} 
at the LHC.


The spontaneous breaking of $U(1)_{B-L}$ is achieved via the first term 
in the second line of Eq.~(\ref{W}), which gives the following potential
for unbroken SUSY
\beq
V=\kappa^2|\Phi\bar{\Phi}-M^2|+\kappa^2|S|^2(|\Phi|^2+|\bar{\Phi}|^2)+
{\rm D-terms}.
\label{V}
\eeq
Here we assumed that the mass parameter $M$ and the dimensionless coupling 
constant $\kappa$ are made real and positive by field rephasing, and the 
scalar components of the superfields are denoted by the same symbol.
Vanishing of the D-terms implies that $|\Phi|=|\bar{\Phi}|$, which yields
$\bar{\Phi}^*=e^{i\varphi}\Phi$, while the F-terms vanish for $S=0$ and 
$\Phi\bar{\Phi}=M^2$, which requires that $\varphi=0$. So, after rotating 
$\Phi$ and $\bar{\Phi}$ to the positive real axis by a $B-L$ transformation, 
we find that the SUSY vacuum lies at
\beq
S=0 \quad {\rm and}\quad \Phi=\bar{\Phi}=M.  
\eeq
The mass spectrum of the scalar $S-\Phi-\bar{\Phi}$ system is constructed
by writing $\Phi=M+\delta\Phi$ and $\bar{\Phi}=M+\delta\bar{\Phi}$. In the 
unbroken SUSY limit, we find two complex scalar fields $S$ and 
$\theta=(\delta\Phi+\delta\bar{\Phi})/\sqrt{2}$ with equal masses 
$m_{S}=m_{\theta}=\sqrt{2}\kappa M$. Soft SUSY breaking can, of course, 
mix these fields and generate a mass splitting. For example, the trilinear
soft term $A\kappa S \Phi\bar{\Phi}$ yields a mass squared splitting 
$\pm\sqrt{2}\kappa M A$ with the mass eigenstates now being 
$(S+\theta^*)/\sqrt{2}$ and $(S-\theta^*)/\sqrt{2}$. This splitting is 
small for $A\ll\sqrt{2}\kappa M$. Let us assume that the mixing 
is in general sub-dominant and ignore it. This simplifies our analysis.

The soft SUSY breaking terms 
\beq
V_1=A\kappa S \Phi\bar{\Phi}-(A-2m_{3/2})\kappa M^2 S
\label{V1}
\eeq
in the potential with $m_{3/2}$ being the gravitino mass and 
$A\sim m_{3/2}$, which arise from the first term in the second line 
of Eq.~(\ref{W}), play an important role in our scheme. Here we 
assume minimal supergravity so that the coefficients of the trilinear 
and linear soft terms are related as shown. Substituting 
$\Phi=\bar{\Phi}=M$ in Eq.~(\ref{V1}), we obtain a linear term in $S$ 
which, together with the mass term $2\kappa^2 M^2 |S|^2$ of $S$, 
generates \cite{Dvali:1994wj} a VEV for $S$:
\beq
\vev{S}=-\frac{m_{3/2}}{\kappa}.
\label{vev}
\eeq 
Substituting this VEV of $S$ in the superpotential
term $\lambda_{\mu}SH_{u}H_{d}$, we obtain \cite{Dvali:1997uq,King:1997ia} 
the MSSM $\mu$ term with $\mu=-\lambda_{\mu}m_{3/2}/\kappa$. The crucial 
point here is that the same VEV generates mass terms 
$-\lambda_{D}m_{3/2}D\bar{D}/\kappa$ for the vector-like superfields $D$, 
$\bar{D}$ via the superpotential terms $\lambda_{D}SD\bar{D}$. The 
trilinear terms corresponding to these superpotential terms
will produce mixing between the scalar components of $D$, $\bar{D}$.
However, we will assume here that this mixing is sub-dominant. 

To preserve gauge coupling unification one should introduce 
vector-like color singlet, $SU(2)_L$ doublet superfields $L$, $\bar{L}$ 
equal in number to the color triplets $D$, $\bar{D}$. (With $D$, 
$\bar{D}$ and $L$, $\bar{L}$ masses $\sim {\rm TeV}$, the gauge 
couplings stay in the perturbative domain for up to four such 
pairs.) These fields with a superpotential coupling 
$\lambda_L S L\bar{L}$ can enhance the branching ratio of 
the decay of the spin zero fields $S$ and $\theta$ to photons and, in 
addition, allow the decay into $W^{\pm}$. 
Introducing the superpotential coupling $Lle^c$, the 
hypercharge of $L$ ($\bar{L}$) is -1/2 (1/2). Their baryon, lepton 
numbers, and R-charges are all zero. These quantum numbers allow the 
superpotential couplings $SLH_u$, $SH_d\bar{L}$, $Lqd^c$, $\bar{L}qu^c$, 
and $\bar{L}l\nu^c$. Substituting $\vev{S}$
in $\lambda_L S L\bar{L}$, the superfields $L$, 
$\bar{L}$ acquire a mass $m_L=-\lambda_L m_{3/2}/\kappa$. 


\begin{figure}[t]
\centerline{\epsfig{file=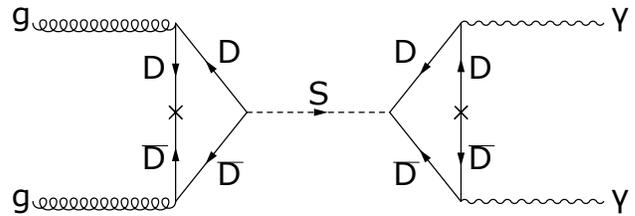,width=8.7cm}}
\caption{
Production of the bosonic component 
of $S$ at the LHC by gluon ($g$) fusion and its subsequent decay 
into photons ($\gamma$). Solid (dashed) lines represent the 
fermionic (bosonic) component of the indicated superfields. The 
arrows depict the chirality of the superfields and the crosses 
are mass insertions which must be inserted in each of the lines
in the loops.}
\label{fig1}
\end{figure}

The real scalar $S_1$ and real pseudoscalar $S_2$ components 
of $S~(=(S_1+iS_2)/\sqrt{2})$ with 
mass $m_S=\sqrt{2}\kappa M$ in the exact SUSY limit can be 
produced at the LHC by gluon fusion via a 
fermionic $D$, $\bar{D}$ loop as indicated in Fig.~\ref{fig1}.
In the absence of the vector-like $L$, $\bar{L}$ superfields,
they can decay into gluons, photons, or $Z$ gauge bosons via 
the same loop diagram, but not to $W^{\pm}$ bosons since the 
$D$, $\bar{D}$ are $SU(2)_L$ singlets. The most 
promising decay channel to search for these resonances is
into two photons with the relevant diagram 
also shown in Fig.~\ref{fig1}. 

Applying the results of S.F.~King and R.~Nevzorov in 
Ref.~\cite{theory}, the cross section of the diphoton 
excess is 
\beq
\label{sigma}
\sigma(pp\rightarrow S_i\rightarrow\gamma\gamma)\simeq 
\frac{C_{gg}}{m_{S}s\Gamma_{S_i}}
\Gamma(S_i\rightarrow gg)\Gamma(S_i\rightarrow \gamma\gamma),
\eeq     
where $i=1,2$, $C_{gg}\simeq 3163$, $\sqrt{s}\simeq 
13~{\rm TeV}$, $\Gamma_{S_i}$ is the total decay width of 
$S_i$, and the decay widths of $S_i$ to two gluons ($g$) or 
two photons ($\gamma$) are given by 
\bea
\label{Sg}
\Gamma(S_i\rightarrow gg)=\frac{n^2\alpha_s^2m_S^3}
{256\,\pi^3\vev{S}^2}\, A_i^2(x),\\
\label{Sgamma}
\Gamma(S_i\rightarrow \gamma\gamma)=\frac{n^2\alpha_Y^2m_S^3
\cos^4\theta_W}{4608\,\pi^3\vev{S}^2}\, A_i^2(x).
\eea 
Here $n$ is the number of $D$, $\bar{D}$ pairs, which are 
taken, for simplicity, to have a common coupling constant $\lambda_D$ 
to $S$, $A_1(x)=2x[1+(1-x)\arcsin^2(1/\sqrt{x})]$, 
$A_2(x)=2x\arcsin^2(1/\sqrt{x})$, $x=4m_D^2/m_S^2> 1$, and 
$\alpha_s$, $\alpha_Y$ are the strong and hypercharge fine-structure 
constants. If the $L$, $\bar{L}$ superfields are present, they will
also contribute to the decay width of $S$ to photons via loop 
diagrams similar to the ones in the right part of Fig.~\ref{fig1}, and 
Eq.~(\ref{Sgamma}) will be replaced by
\bea
\label{SgammaL}
\Gamma(S_i\rightarrow \gamma\gamma)&=&\frac{n^2m_S^3\alpha_Y^2\cos^4\theta_W}
{4608\,\pi^3\vev{S}^2}A_i^2(x)
\nonumber\\
& &\left[1+\frac{3A_i(y)}{2A_i(x)}\left(1+\frac{\alpha_2\tan^2\theta_W}
{\alpha_Y}\right)\right]^2,
\eea 
where $\alpha_2$ is the $SU(2)_L$ fine-structure constant and 
$y=4m_L^2/m_S^2> 1$.

The cross section in Eq.~(\ref{sigma}) simplifies under the 
assumption that the spin zero fields $S_i$ decay predominantly into 
gluons, namely, $\Gamma_{S_i}\simeq\Gamma(S_i\rightarrow gg)$. In 
this case, as pointed by R.~Franceschini et al. 
in Ref.~\cite{theory}, one obtains $\sigma(pp\rightarrow S_i
\rightarrow\gamma\gamma)\simeq 8~{\rm fb}$ if
\beq
\label{condition}
\frac{\Gamma(S_i\rightarrow \gamma\gamma)}{m_{S}}
\simeq 1.1 \times 10^{-6}.
\eeq
For $x$ and $y$ just above unity, which guarantees that the 
decay of $S_i$ to $D$, $\bar{D}$ and $L$, $\bar{L}$ pairs is 
kinematically blocked, $A_1(x)$ and $A_2(y)$ are 
maximized with values $A_1\simeq 2$ and $A_2\simeq \pi^2/2$.
Note that $x$ close to unity means $m_D\simeq 375~{\rm GeV}$. 
However, one should work with somewhat larger $m_D$ as indicated 
by ATLAS and CMS. So we take $m_D=700~{\rm GeV}$. It is more 
beneficial to consider the decay of the pseudoscalar $S_2$ since 
$A_2(x)>A_1(x)$ for all $x>1$. Using Eq.~(\ref{SgammaL}), we 
then find that the condition in Eq.~(\ref{condition}) is 
satisfied for $n m_S/|\vev{S}|\simeq 2.97$. Therefore, for $n=3$, 
we require that $m_S/|\vev{S}|\simeq 0.99$, which, for $m_S\simeq 
750~{\rm GeV}$, implies that $|\vev{S}|\simeq 758~{\rm GeV}$. 
In this case, $\lambda_D\simeq 0.92$ and $\lambda_L$ is just 
above 0.49. Comparing Eqs.~(\ref{Sgamma}) and (\ref{SgammaL}), 
we find that the inclusion of the vector-like $L$, $\bar{L}$ 
superfields enhances the decay width of $S_2$ to photons by 
about a factor 58.5. 


In the exact SUSY limit, the complex scalar field $S$ could decay 
into MSSM Higgsinos (potential dark matter candidate) via the 
superpotential term 
$\lambda_{\mu}SH_{u}H_{d}$ if this is kinematically allowed. 
$S$ also could decay into right handed 
sneutrinos via the F-term $F_{\bar{\Phi}}$ between the superpotential 
terms $\kappa S\Phi\bar{\Phi}$ and $\bar{\Phi}\nu^{c}\nu^{c}$ after 
substituting the VEV of $\Phi$. 
The decay 
widths in the two cases are
\beq
\Gamma^S_H=\frac{\lambda_{\mu}^2}{8\pi}m_{S}, 
\quad 
\Gamma^S_{\nu^c}=\frac{\lambda_{\nu^c}^2}{8\pi}m_{S},
\label{decay}
\eeq 
respectively, where we assumed that the masses of the Higgsinos and 
the relevant right handed sneutrinos are much smaller than $m_S$.
Depending on the kinematics the total decay width of the 
resonance could easily lie in the multi-GeV range.
The diphoton, dijet, and diboson decay modes in this case
would be sub-dominant. 

Our estimate of $m_S/|\vev{S}|$ 
after Eq.~(\ref{condition}) requires that the decay widths of
$S$ into MSSM Higgsinos and right handed sneutrinos are sub-dominant 
or kinematically blocked. The latter is achieved for $|\mu|=
\lambda_{\mu}|\vev{S}|>m_S/2\simeq 375~{\rm GeV}$ (or 
$\lambda_\mu\gtrsim 0.49$) and $\lambda_{\nu^c}M>m_S/2$. 
Demanding that the mass of the $B-L$ gauge boson $m_{Z'}=
\sqrt{6}g_{B-L}M>3~{\rm TeV}$ \cite{Zprime,okada-okada}, say, we 
find that $g_{B-L}M\gtrsim 1225~{\rm GeV}$ ($g_{B-L}$ is the GUT 
normalized $B-L$ gauge coupling constant). From 
Eq.~(\ref{vev}) and setting, say, 
$m_{3/2}=50~{\rm GeV}$, we obtain
$\kappa\simeq 0.066$, $M\simeq 8040~{\rm GeV}$, 
$\lambda_{\nu^c}\gtrsim 0.047$, and $g_{B-L}\gtrsim 0.15$. 
A gravitino in this mass range is a plausible cold matter 
candidate -- for a recent discussion and references, see 
Ref.~\cite{Okada:2015vka}. Finally, we have checked that, in this 
example, $g_{B-L}\lesssim 0.25$, $\lambda_{D}$, and $\lambda_{L}$ 
remain perturbative up to the GUT scale ($M_{\rm GUT}$). In 
particular, if $g_{B-L}\simeq 0.24$, it unifies with the MSSM 
gauge coupling constants at $M_{\rm GUT}$. So the requirements 
for a viable diphoton resonance are met.   

\begin{figure}[t]
\centerline{\epsfig{file=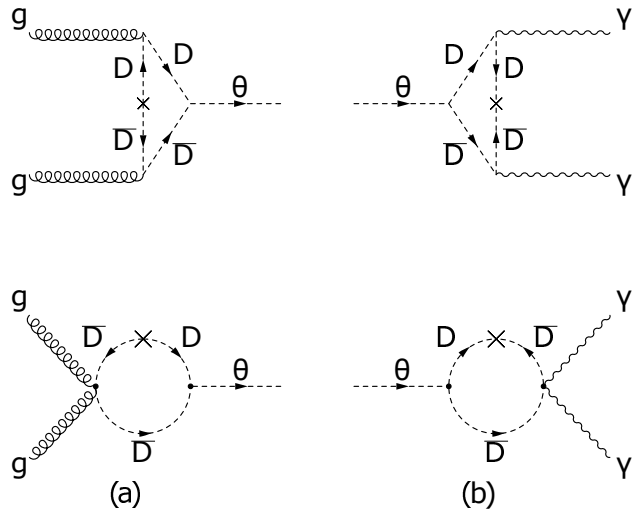,width=8.7cm}}
\caption{
Production of the bosonic component 
of $\theta$ at the LHC by gluon ($g$) fusion and its subsequent 
decay into photons ($\gamma$). The notation is the same as in 
Fig.~\ref{fig1} with the crosses indicating mass squared 
insertions.
}
\label{fig3}
\end{figure}

The complex spin zero field $\theta=(\delta\Phi+\delta\bar{\Phi})/
\sqrt{2}=(\theta_1+i\theta_2)/\sqrt{2}$, which consists of a 
real scalar ($\theta_1$) and a real pseudoscalar ($\theta_2$) field
and has mass $m_{\theta}=\sqrt{2}\kappa M$ in the SUSY limit, 
couples to the scalar vector-like fields $D$, $\bar{D}$ via
the F-term $F_S$ between the superpotential terms 
$\kappa S\Phi\bar{\Phi}$ and $\lambda_D S D\bar{D}$. The 
coupling constant is $\lambda_D m_{\theta}$. It also can be produced
at the LHC by gluon fusion via scalar $D$, $\bar{D}$ loops as shown 
in Fig.~\ref{fig3}(a), and decay into two photons via the diagrams in 
Fig.~\ref{fig3}(b). (In the presence of $L$, $\bar{L}$ superfields 
similar diagrams with scalar $L$, $\bar{L}$ loops also contribute 
to the decay of $\theta$ into photons.) The important point here is 
that the mass squared 
insertions in all the diagrams of Fig.~\ref{fig3} arise from the soft SUSY 
breaking trilinear term $A^\prime\lambda_D S D\bar{D}$ and are thus 
equal to $A^\prime m_{D}$, where $m_{D}=-\lambda_{D}m_{3/2}/\kappa$ is 
the mass of the $D$, $\bar{D}$ superfields generated by the VEV of $S$
in Eq.~(\ref{vev}). Consequently, for $A^\prime\ll m_{D}$, the cross 
sections for the diphoton excess are suppressed by a factor
$(A^\prime/m_{D})^4$ relative to the ones for the spin zero field $S$.
Larger soft SUSY breaking trilinear terms will
enhance the diagrams in Fig.~\ref{fig3} and also cause larger mixing 
between $S$ and $\theta$. In this case all four spin zero states can 
contribute to the diphoton excess. 
          
The field $\theta$ can decay, in the exact SUSY
limit, into MSSM Higgs fields via the F-term $F_S$ between the 
superpotential terms $\kappa S \Phi\bar{\Phi}$ and $\lambda_{\mu} S 
H_{u}H_{d}$ if this is kinematically allowed. 
The relevant coupling constant is $\lambda_{\mu}m_{\theta}$, 
and thus the decay width is the same as $\Gamma^S_H$ in 
Eq.~(\ref{decay}) provided that the masses of the Higgs fields are much 
smaller than $m_{\theta}$ (for $m_S=m_{\theta}$). $\theta$ 
also could decay into right handed neutrinos via the superpotential term 
$\lambda_{\nu^c}\bar{\Phi}\nu^c\nu^c$ 
with a decay 
width equal to $\Gamma^S_{\nu^c}$ in Eq.~(\ref{decay}) under the same 
assumption.



In conclusion, we have presented a realistic SUSY model based on a 
$U(1)_{B-L}$ extension 
of the MSSM that contains resonances observable at the LHC and/or future 
colliders. The underlying gauge and R-symmetries are such that the 
MSSM $\mu$ parameter and the masses of vector-like 
superfields and a gauge singlet superfield cannot 
be arbitrarily large. A resonance system consisting of four spin 
zero states arises from a 
gauge singlet scalar and a pair of conjugate Higgs superfields 
responsible 
for the $B-L$ breaking. These states are degenerate in mass in the SUSY 
limit, and depending on the details of SUSY breaking, one or more of 
these states could explain the observed 750 GeV diphoton excess. Their 
total decay widths can lie in 
the multi-GeV range, depending on the kinematics, in which 
case the diphoton, diboson and dijet events will be sub-dominant. 

\acknowledgments{Q.S. is supported in part by the DOE grant 
DOE-SC0013880. We thank George Leontaris for discussions and Aditya 
Hebbar for help with the figures.}

\def\ijmp#1#2#3{{Int. Jour. Mod. Phys.}
{\bf #1},~#3~(#2)}
\def\plb#1#2#3{{Phys. Lett. B }{\bf #1},~#3~(#2)}
\def\zpc#1#2#3{{Z. Phys. C }{\bf #1},~#3~(#2)}
\def\prl#1#2#3{{Phys. Rev. Lett.}
{\bf #1},~#3~(#2)}
\def\rmp#1#2#3{{Rev. Mod. Phys.}
{\bf #1},~#3~(#2)}
\def\prep#1#2#3{{Phys. Rep. }{\bf #1},~#3~(#2)}
\def\prd#1#2#3{{Phys. Rev. D }{\bf #1},~#3~(#2)}
\def\npb#1#2#3{{Nucl. Phys. }{\bf B#1},~#3~(#2)}
\def\np#1#2#3{{Nucl. Phys. B }{\bf #1},~#3~(#2)}
\def\npps#1#2#3{{Nucl. Phys. B (Proc. Sup.)}
{\bf #1},~#3~(#2)}
\def\mpl#1#2#3{{Mod. Phys. Lett.}
{\bf #1},~#3~(#2)}
\def\arnps#1#2#3{{Annu. Rev. Nucl. Part. Sci.}
{\bf #1},~#3~(#2)}
\def\sjnp#1#2#3{{Sov. J. Nucl. Phys.}
{\bf #1},~#3~(#2)}
\def\jetp#1#2#3{{JETP Lett. }{\bf #1},~#3~(#2)}
\def\app#1#2#3{{Acta Phys. Polon.}
{\bf #1},~#3~(#2)}
\def\rnc#1#2#3{{Riv. Nuovo Cim.}
{\bf #1},~#3~(#2)}
\def\ap#1#2#3{{Ann. Phys. }{\bf #1},~#3~(#2)}
\def\ptp#1#2#3{{Prog. Theor. Phys.}
{\bf #1},~#3~(#2)}
\def\apjl#1#2#3{{Astrophys. J. Lett.}
{\bf #1},~#3~(#2)}
\def\apjs#1#2#3{{Astrophys. J. Suppl.}
{\bf #1},~#3~(#2)}
\def\n#1#2#3{{Nature }{\bf #1},~#3~(#2)}
\def\apj#1#2#3{{Astrophys. J.}
{\bf #1},~#3~(#2)}
\def\anj#1#2#3{{Astron. J. }{\bf #1},~#3~(#2)}
\def\mnras#1#2#3{{MNRAS }{\bf #1},~#3~(#2)}
\def\grg#1#2#3{{Gen. Rel. Grav.}
{\bf #1},~#3~(#2)}
\def\s#1#2#3{{Science }{\bf #1},~#3~(#2)}
\def\baas#1#2#3{{Bull. Am. Astron. Soc.}
{\bf #1},~#3~(#2)}
\def\ibid#1#2#3{{\it ibid. }{\bf #1},~#3~(#2)}
\def\cpc#1#2#3{{Comput. Phys. Commun.}
{\bf #1},~#3~(#2)}
\def\astp#1#2#3{{Astropart. Phys.}
{\bf #1},~#3~(#2)}
\def\epjc#1#2#3{{Eur. Phys. J. C}
{\bf #1},~#3~(#2)}
\def\nima#1#2#3{{Nucl. Instrum. Meth. A}
{\bf #1},~#3~(#2)}
\def\jhep#1#2#3{{J. High Energy Phys.}
{\bf #1},~#3~(#2)}
\def\jcap#1#2#3{{J. Cosmol. Astropart. Phys.}
{\bf #1},~#3~(#2)}
\def\lnp#1#2#3{{Lect. Notes Phys.}
{\bf #1},~#3~(#2)}
\def\jpcs#1#2#3{{J. Phys. Conf. Ser.}
{\bf #1},~#3~(#2)}
\def\aap#1#2#3{{Astron. Astrophys.}
{\bf #1},~#3~(#2)}
\def\mpla#1#2#3{{Mod. Phys. Lett. A}
{\bf #1},~#3~(#2)}

\end{document}